\documentclass[%reprint,
reprint, %% uncomment this for 2 columns
preprintnumbers,
 amsmath,amssymb,
 aps,
prl,
]{revtex4-2}

\def\isotope#1#2{\mbox{${}^{#2}{\rm #1}$}}
\def\fe6#1{\isotope{Fe}{6#1}}
\def\pu24#1{\isotope{Pu}{24#1}}
\def\I12#1{\isotope{I}{12#1}}
\def\hf18#1{\isotope{Hf}{18#1}}
\def\cm24#1{\isotope{Cm}{24#1}}
\def\mature{I_s/{\rm FeO}}

\usepackage{graphicx}
\usepackage{dcolumn}
\usepackage{bm}
\usepackage{hyperref}
\hypersetup{
    colorlinks,
    linkcolor={red!80!black},
    citecolor={red!80!black},
    urlcolor={blue!80!black}
}
\usepackage{color}
\usepackage{orcidlink}

\begin{document}

\preprint{KCL-PH-TH/2026-10, CERN-TH-2026-073}

\title{Gardening on the Moon: An Advection-Diffusion Model to Guide the Search for Supernova Debris in the Lunar Regolith}

\author{Emily~S. Costello
\orcidlink{0000-0001-7939-9867}}
\email{costello@higp.hawaii.edu}
\affiliation{Hawai\textquoteleft i Institute of Geophysics and Planetology, University of Hawai\textquoteleft i at Mānoa, Honolulu, HI~96822, USA}

\author{John~Ellis \orcidlink{0000-0002-7399-0813}}
\affiliation{Theoretical Particle Physics and Cosmology Group, Department of Physics, King’s College London, London WC2R 2LS, UK; \\
Theoretical Physics Department, CERN, CH-1211 Geneva 23, Switzerland}

\author{Brian~D.~Fields \orcidlink{0000-0002-4188-7141}}
\affiliation{Department of Astronomy, University of Illinois, Urbana, IL 61801, USA}

\author{Rebecca~Surman \orcidlink{0000-0002-4729-8823}}
\affiliation{Department of Physics and Astronomy, University of Notre Dame, Notre Dame, IN 46556, USA}

\author{Xilu Wang
\orcidlink{0000-0002-5901-9879}}
\affiliation{State Key Laboratory of Particle Astrophysics, Institute of High Energy Physics, Chinese Academy of Sciences, Beijing 100049, China}

\begin{abstract}
The vertical redistribution of materials in the lunar regolith—ranging from continuously produced space-weathering products to sporadic pulses of supernova- or kilonova-derived isotopes—remains a fundamental problem in planetary science. We present a unified stochastic model of regolith gardening induced by the impact flux. Treating gardening as a competition between impact-driven advection and diffusion predicts the maturity profiles of {\it Apollo} cores over more than two orders of magnitude in time ($1.4 \times 10^7$ to $4.5 \times 10^8$ years). This model describes well the depth profiles of live \fe60 in {\it Apollo} regolith samples, suggesting that supernova dust capture is independent of native iron abundance, and is consistent with a uniform influx at the latitudes of the {\it Apollo} landing sites. We extend our model to predict lunar signals for live {\em r}-process species that might originate from supernovae or kilonovae: \pu244 tied to terrestrial detections, and \I129, \hf182, and \cm247 based on {\em r}-process calculations. The \pu244/\fe60 depth profile can probe the origin of \pu244, motivating searches in {\it Artemis} regolith samples down to depths ${\cal O}(100)$~cm.
\end{abstract}

\maketitle

\section{Introduction}

The lunar regolith serves as a record-keeper of the Solar System’s history, preserving evidence of solar wind, cosmic rays, and episodic interstellar events. However, this record is subject to scrambling by impact gardening. Gardening is a stochastic vertical transport process driven by the flux of crater-forming impacts over twelve orders of magnitude in scale \cite{gault1974mixing,costello2018mixing}. Classically, the cumulative effects of gardening are treated as a diffusive `random walk' (see, e.g.,~\cite{siegler2016lunar}). However, high-resolution depth profiles of regolith maturity (the optical and physical changes that result from exposure to space) and isotopic tracers consistently exhibit sharp gradients and vertical centroids that are mathematically incompatible with purely diffusive smearing (see, e.g.,~\cite{morris1978situ}).

{Modeling the gardening of the regolith is crucial for studying} the infall of debris from nearby astrophysical events.
It was suggested in~\cite{Ellis:1995qb} that nearby supernova explosions might have deposited on Earth detectable amounts of \fe60 and other live isotopes such as \pu244 (for a recent review see \cite{Fields2023}). Deep-sea deposits of \fe60 were subsequently reported by~\cite{Knie:1999zz} and many subsequent experiments, and \fe60 has also been found in {\it Apollo} samples of the lunar regolith~\cite{fimiani2016interstellar}. Subsequently, \pu244 has also been discovered in deep-ocean deposits~\cite{Wallner2021}. The terrestrial \fe60 was mainly deposited in a pulse about 2.3~Mya~(million years ago) \cite{Ertel2023}, 
{with a smaller,}
earlier pulse about 7.3~Mya~\cite{Wallner2021}. {These pulses could be due either to shock fronts passing through the Solar System or to the Earth passing through overdensities in the local interstellar medium due to clouds or walls of the bubble blown by prior supernova explosions.}  
\pu244 has also been detected in deep-sea sample; the signal 
span both pulses,
but the earlier time structure is unclear~\cite{Wallner2021}. This has provoked discussion~\cite{Wang2023} whether or not the \fe60 and \pu244 have a common origin. The production of \fe60 is expected in core-collapse supernovae, whereas \pu244 would have been produced by the astrophysical {\em r}-process, which may take place in exceptional supernovae or in kilonovae (collisions of neutron stars)~\cite{LIGOScientific:2017ync}, with the latter likely to produce \pu244 more copiously, {along with iodine and other elements that are essential for human life~\cite{Ellis:2024uvj}}. Since \pu244 has a much longer half-life ($t_{1/2} = 81.3$ Myr) than \fe60 ($t_{1/2} = 2.60(5)$ Myr), deposits over much longer periods may be detectable, and their time structure could cast light on their origin, as could those of other {\em r}-process isotopes such as \I129, \hf182, and \cm247~\cite{Wang2023}.

{Conversely}, supernova-produced \fe60 may serve as a critical new probe of vertical transport mechanisms in the regolith.
Isotopic analysis of {\it Apollo} 11, 12, 15 and 16 samples revealed \fe60 activities exceeding cosmic-ray production by a factor ${\cal O}(30)$~\cite{fimiani2016interstellar,Zwickel2026}. While this signal likely records the deposition of supernova debris around 2 million years ago, the present depth profile is a convolution of the initial deposition flux and the subsequent mechanical gardening, {as we now discuss}.

\section{Gardening Model}

Every impact crater on the Moon redistributes regolith: excavating, compacting, and burying material under ejecta. We present here a mass-conservative analytic model of the transport efficiency of these processes, updating previous frameworks~\cite{gault1974mixing, costello2018mixing,costello2020impact,costello2021secondary}. Defining depth $z$ as positively downward from the vacuum-regolith interface ($z=0$) and $t$ as the accumulated duration of exposure, we express the net vertical transport depth $D_z(t)$ as the sum of downward drivers (burial and compaction) minus the upward driver (exhumation):
\begin{equation}
\label{eq:Dz_Full}
\begin{aligned}
D_z(t) & = D_B(t) + D_C(t) - D_E(t) \\
& %\hspace{-1cm} = 
\underbrace{ \left[ \eta_b(b) + \frac{h_c}{c_c^{-\frac{2}{b-2}}} - \frac{h_e}{c_e^{-\frac{2}{b-2}}} \right] }_{\text{Geometric Efficiency}} \underbrace{ \left( \frac{\pi a b t}{4(b-2)\lambda} \right)^{\frac{1}{b-2}} }_{\text{Probabilistic Frequency}} \, 
\\
\end{aligned}
\end{equation}
where $\eta_b(b)$ is the effective burial efficiency (defined as a summation over ejecta annuli), $h_c$ and $h_e$ are the vertical reach scales for compaction and excavation, $c_c$ and $c_e$ are their respective horizontal interaction scales, and $a, b,$ and $\lambda$ characterize the impact production function and frequency (see Supplement S1 for full derivations and parameter values).

We derive the vertical transport velocity $v_z(t)$ by differentiating Eq.~(\ref{eq:Dz_Full}) with respect to $t$. Because $D_z \propto t^{1/(b-2)}$, 
we have
\begin{equation}
v_{z}(t) = \frac{\partial D_{z}}{\partial t} = 
\frac{D_{z}(t)}{(b-2)t}
\label{eq:velocity}
\end{equation}
\noindent This derivative defines the direction of transport relative to the impact population. For steep secondary power-law slopes ($b > 2$, which we will adopt), the factor $(b-2)$ is positive, yielding $v_z > 0$. 
In our coordinate system, this corresponds to the net downward advection of surface-delivered materials. Conversely, for shallower primary slopes ($b < 2$), $v_z < 0$, representing the upward exhumation  of material toward the surface.

Stochastic vertical mixing by numerous smaller craters above this advective front is captured by the diffusion coefficient $\kappa(t)$, defined as a function of the characteristic mixing length $z_\kappa \approx 0.54 D_z$:

\begin{equation}
\kappa(t) = \frac{z_{\kappa}^{2}}{t} = \left(\frac{\Gamma}{H}\right)^{2} \frac{D_{z}(t)^{2}}{t}
\label{eq:diffusion}
\end{equation}
\noindent 
where $\Gamma \approx 0.11$ is a dimensionless gardening efficiency determined by the variance of impact-driven displacements, and $H = 0.2$ is the dimensionless total vertical reach of a transient crater (see Supplement S2).

This kinematic model allows the regolith to be treated as a continuum where the evolution of a concentration or number density $C(z,t)$ is governed by the following Advection-Diffusion Equation (ADE):
\begin{equation}
\frac{\partial C}{\partial t} = \frac{\partial}{\partial z}\left(\kappa(t)\frac{\partial C}{\partial z}\right) - v_z(t)\frac{\partial C}{\partial z} - \lambda_{\text{decay}}C + S(z,t) 
\label{eq:ADE}
\end{equation}
where $S(z,t)$ represents the spatiotemporal source term (continuous for maturity, episodic for supernova-derived \fe60, and continuous/depth-dependent for cosmogenic \fe60) and $\lambda_{\text{decay}}=1/\tau$ is the radioactive decay rate given the mean life $\tau$. When solving this equation, we treat the surface as a flux boundary.

\section{Results}

\subsection{Validation: Regolith Maturity}
\begin{figure}[b]
\includegraphics[width=0.95\linewidth]{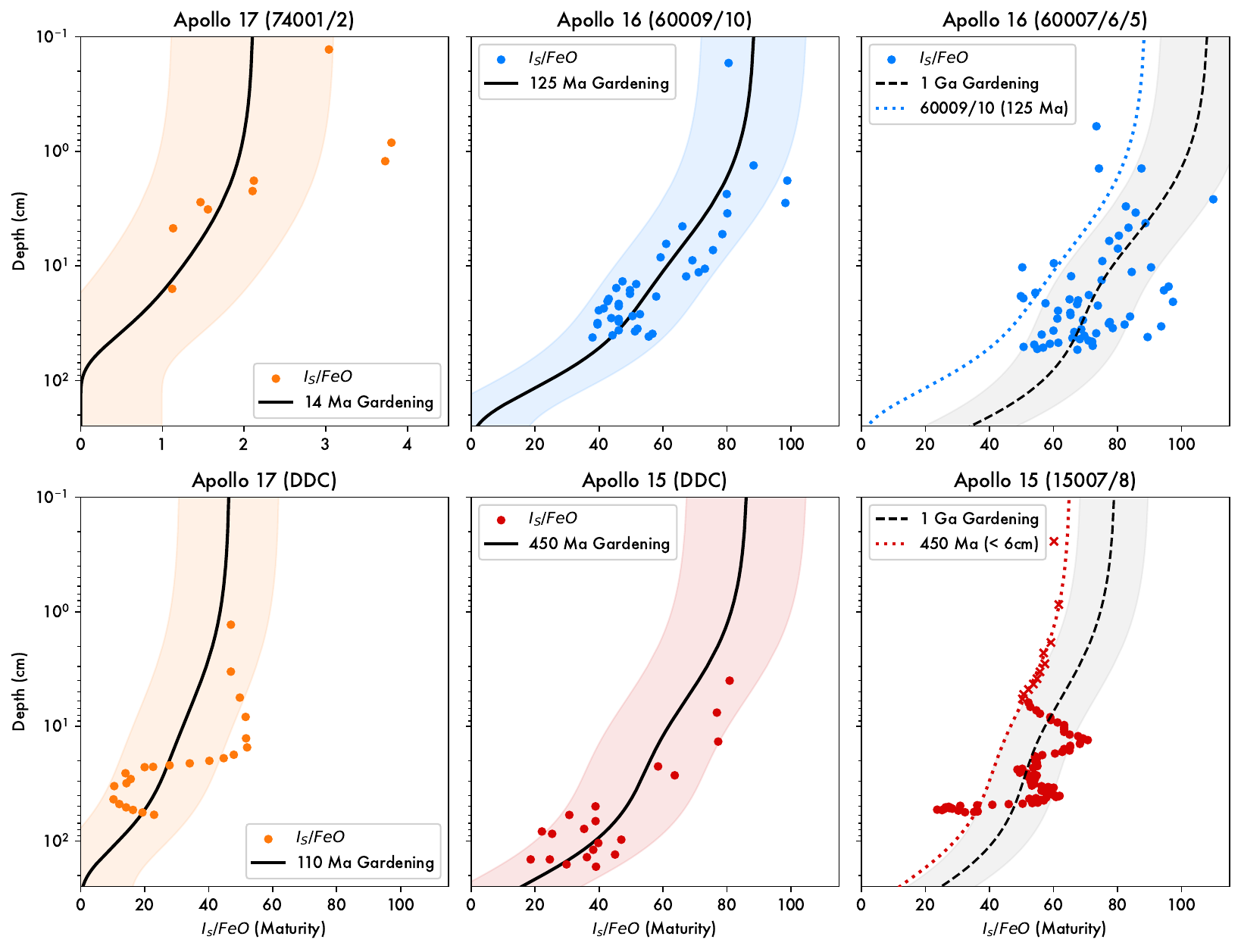}
\caption{\textbf{Surface maturity profiles for {\it Apollo} 15, 16, and 17 samples that validate the unified gardening model.} Observed $\mature$ maturity indices \cite{morris1978situ} (colored points) are compared with gardening model results (black lines). For the four cores on the left, model timescales ($t$) are constrained by independent cosmic ray track and cosmogenic radionuclide densities reported in \cite{morris1978situ}; for the two cores on the right \cite{fimiani2016interstellar}, a 1~Ga reference mixing profile and the gardening profile of the dated core from the same landing region are shown (black dashed and colored dotted lines, respectively). Shaded bands represent the empirical error ($\sigma = 0.24$). The model translates time-integrated gardening flux $\mathcal{M}(z, t)$ into maturity by accounting for the depth-dependent bulk density $\rho(z)$ ($1.3$ to $1.9$~g/cm$^3$; see supplement S3), such that $\mature(z) \propto \mathcal{M}(z, t) / \rho(z)$. In order to account for mineralogy-dependent variations in space weathering efficiency and initial iron abundance between landing sites, model profiles are fluence-matched to the observed data.
\label{fig:Morris} }
\end{figure}

By incorporating the flux of secondary impacts as the primary driver of vertical transport \cite{costello2021secondary}, the model accurately reproduces the characteristic $\mature$ measure of soil maturity (where $I_s$ is the concentration of nanophase metallic iron) versus depth profiles observed across the {\it Apollo} 15, 16, and 17 cores whose ages are constrained by independent cosmic ray track and cosmogenic radionuclide densities that are used as standard benchmarks for regolith gardening~\cite{morris1978situ,heiken1991lunar}, as seen in Fig.~\ref{fig:Morris}. 
This correspondence across over two orders of magnitude in exposure age ($\sim 1.4 \times 10^7$ to $\sim 4.5 \times 10^8$ years) suggests that the observed maturation front is a direct manifestation of the advective-diffusive equilibrium established by the impact size-frequency distribution. {The agreement between the observed depth distribution of $\mature$ in the {\it Apollo} 16 and {\it Apollo} 15 cores analyzed for \fe60 and our model simulation demonstrates that it describes successfully the gardening history of these cores over ages ranging from tens of millions to several hundred million years; sufficiently ancient to have reached a steady-state of cosmogenic isotope production and decay (Fig.~\ref{fig:Morris}).}

While the ADE solution captures bulk transport, the stochastic nature of individual impacts introduces concentration fluctuations, as observed in core data~\cite{morris1978situ} (see Fig.~\ref{fig:Morris}). We utilize these residuals from the maturity validation to define a conservative site-agnostic empirical uncertainty envelope. By aggregating the normalized variance between our most-probable model lines and the observed $\mature$ profiles, we derive a global standard deviation ($\sigma_{global} \approx 0.24$) that represents the inherent noise of the gardening process. This envelope, detailed in S4 of the Supplemental Material, provides a statistically realistic benchmark for the forward-modeled distributions of episodic species such as \fe60 and \pu244, accounting for the fact that any single core is a discrete realization of a locally unique impact history.

\subsection{Episodic Sources: Supernova \fe60}

To evaluate the vertical profile of interstellar debris incident on the Moon, we introduce a spatiotemporal source term $S(z,t)$ driven by a surface flux $\Phi(t)$.
For the case of \fe60, deep-sea measurements show distinct pulses, which ultimately are of supernova origin \cite{Fry2015}.
We thus model the \fe60 flux as a superposition of $n$ Gaussian pulses:
\begin{equation}
\label{eq:source}
S(z,t) = \delta(z)\Phi(t) = \delta(z) \sum_{n} \frac{\mathcal{F}_n}{\sigma_n \sqrt{2\pi}} \exp\left( -\frac{(t - t_{n})^2}{2\sigma_n^2} \right)
\end{equation}
where $\mathcal{F}_n$ is the fluence (time-integrated flux, i.e., number of atoms deposited per unit area) from pulse $n$. Here, the $\delta(z)$ factor dictates that the supernova material falls only on the surface ($z=0$).

As discussed in the Supplemental Material, data on deep-ocean deposits of \fe60 ($t_{1/2} = 2.26$ Myr) provide evidence of two pulses: a stronger pulse that reached Earth about 2.3 million years ago and a weaker pulse that arrived about 7.3 million years ago. It is plausible that there were earlier nearby supernovae~\cite{breitschwerdt2016locations}, but the data on deep-ocean deposits do not extend beyond $\sim 10$ million years ago and, in view of the short half-life of \fe60, an earlier \fe60 pulse {surviving strongly to the present} seems unlikely.

{Pulse parameters appear in Table \ref{tab:flux_params}. To find the ``interstellar'' \fe60 fluence at Earth, we correct the deep-sea crust measurements for the $U_{\rm Fe} = 17\%$ incorporation or uptake efficiency.  Also, we assume the radioisotope flux is isotropic,
which is conservative and is motivated by the ``pinball'' model of magnetically confined dust grains that deliver the radioisotopes \cite{Fry2020}.  Had we followed earlier work and instead assumed that the dust comes from a single direction, then the interstellar flux would be higher by a factor of 4, which corrects for the Earth's surface area to cross section ratio.} 

\begin{figure}[t]
\includegraphics[width=\linewidth]{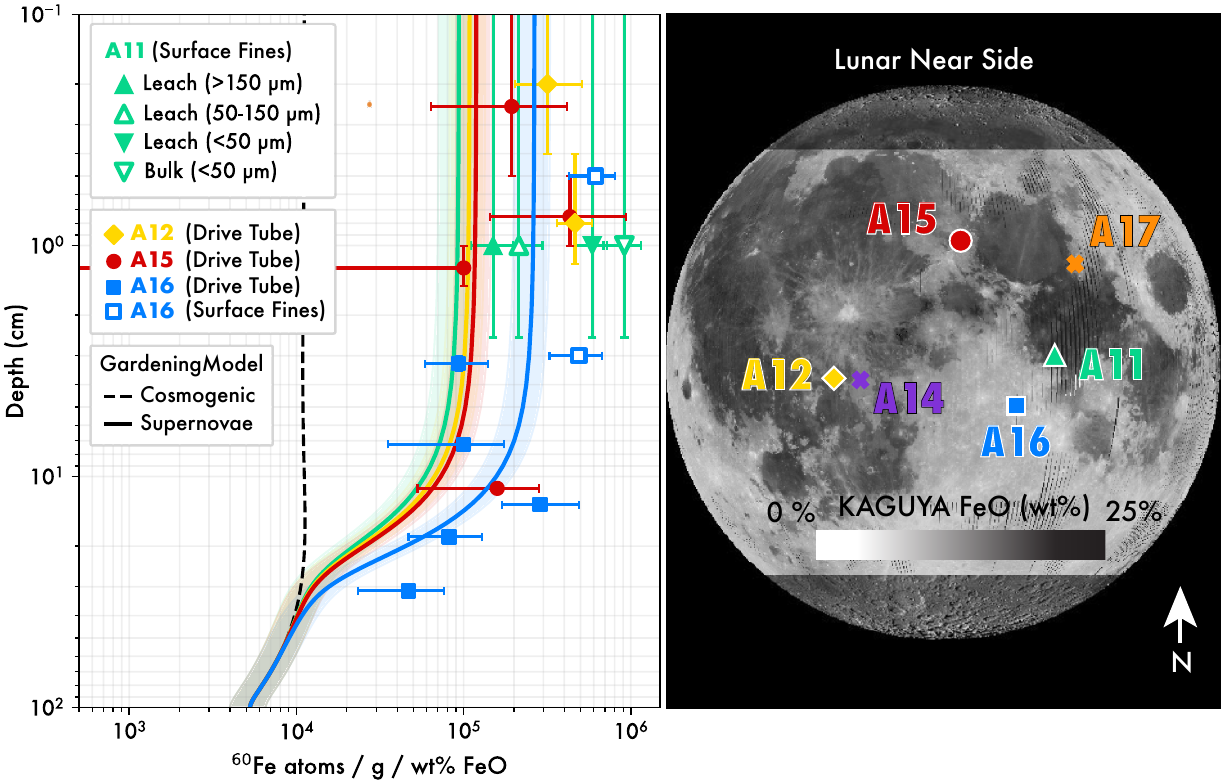}
\caption{\label{fig:60Fe} Left: Depth profile of $^{60}$Fe across {\it Apollo} landing sites. ADE model results for the combination of two known supernova (SN) pulse events are shown as colored solid lines scaled for specific {\it Apollo} sites, compared with lunar soil data from \cite{fimiani2016interstellar} and \cite{Zwickel2026}. To isolate the SN-source signal, measured activities are converted to atoms per unit mass, normalized by the stable Ni concentration and, importantly, normalized by the average FeO wt\% of the respective landing site {(as detailed in Supplement S3.4)}. A11 data~\cite{Zwickel2026} separate bulk dissolved fractions from leached fractions across varying grain sizes. Horizontal error bars represent the propagation of $1\sigma$ measurement uncertainties. The shaded bands around the model contours denote the model uncertainty ($\sigma=0.24$) propagated from the validated gardening parameters in Fig.~\ref{fig:Morris}. The dashed black line represents the modeled steady-state in-situ cosmogenic background floor. Right: The nearside of the Moon, with locations of the {\it Apollo} landing sites shown and a Kaguya Multiband Imager FeO wt\% map overlay \cite{lemelin2015lunar}.}
\end{figure}

The left panel of Fig.~\ref{fig:60Fe} compares lunar measurements in {\it Apollo} 11, 12, 15 and 16 samples of the excess \fe60 (after subtraction of the cosmic-ray contribution) with the prediction of the ADE model with the two known \fe60 pulses (lines with uncertainty shaded). We see that, despite having the different native iron contents~\cite{TRANG2019307,lemelin2015lunar} shown in the right panel of Fig.~\ref{fig:60Fe}, the {\it Apollo} 15 and 16 core samples show similar \fe60 inventories that are consistent with the ADE model predictions.  
This agreement boosts confidence in the ADE gardening model, and also shows that the adopted \fe60 flux is consistent with the lunar data.  This suggests that the radioisotope flux may be closer to isotropic than uni-directional.

{The finest fraction ($<50\mu$m) of mature surface regolith from A11 \cite{Zwickel2026}, and samples of the upper few centimeters of soil from A16 and the A12 drive tube \cite{fimiani2016interstellar}, exhibit higher \fe60 concentrations than predicted by models of cosmogenic production, supernova delivery, and impact gardening (Fig.~\ref{fig:60Fe}). This anomaly suggests that the smallest, most mature size fractions of lunar soil are exceptionally efficient at capturing exogenic \fe60 while simultaneously suppressing the cosmogenic background. This dual effect is likely governed by space weathering, as discussed in S5.}

As was pointed out in~\cite{Fry2016}, if propagation of the supernova debris containing \fe60 retains directional information until its arrival in the Solar System, the dependence of density of \fe60 deposition on lunar latitude could provide insight into the location of the source of the \fe60. The difference in latitude between the {\it Apollo} 15 and 16 landing sites (see the right panel of Fig.~\ref{fig:60Fe}) provides a limited lever arm, and it is not surprising that the {\it Apollo} 15 and 16 samples have similar \fe60 inventories. {\it Artemis}
measurements of the \fe60 inventory at the lunar South Pole could provide valuable insight. However, it is possible that the directional information may have been lost during propagation due to deflection of \fe60-carrying dust grains by interstellar magnetic fields~\cite{Fry2020,Ertel2023}.

\begin{figure}[t]
\includegraphics[width=\linewidth]{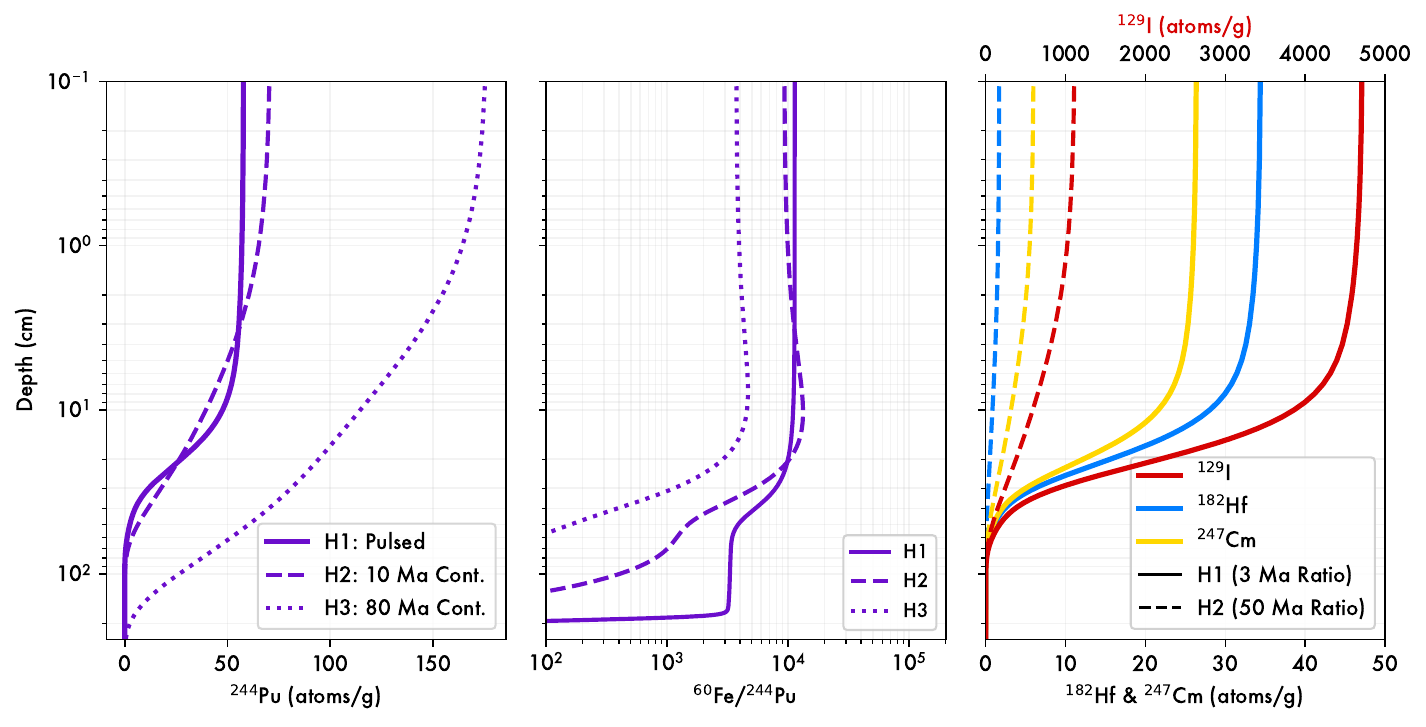}
\caption{\label{fig:244Pu} {\bf Left panel}: Forward modeled \pu244 depth profiles in various delivery histories. The unified gardening model is applied to {three} depositional histories: recent pulsed delivery (H1; solid), and {two} continuous-flux scenarios ranging from short-term accumulation (H2, $10$~Myr; dashed) to long-term  steady-state saturation (H3, $80$~Myr; dotted). All profiles are normalized by the mass-density $\rho(z)$ (gray dashed line, top axis, supplement S6) to present abundances in atoms per unit mass. We see that depth-resolved \pu244 profiles can distinguish between {recent astrophysical events  and a longer downpour of interstellar dust} {but give no hints as to whether the \pu244 and \fe60 sources are the same or distinct}.
{\bf Center panel}: The ratio of \fe60 and \pu244 abundances as a function of depth in histories H1, H2, H3 for \pu244 production. We see that depth-resolved \pu244 profiles can distinguish clearly between histories H1/H2 and history H3.
{\bf Right Panel}: Depth-integrated {\em r}-process concentrations for $^{129}$I (top axis), and $^{182}$Hf and $^{247}$Cm (bottom axis) following a pulsed delivery centered at 2.3 Mya and 7.3 Mya for H1 (solid lines), and a continuous influx over the last 10 Myr for H2 (dashed lines) with an injection time of 50 Myr. 
The corresponding {\em r}-process isotopes production ratios relative to \pu244 are given in Table~5 and Table~6 of~\cite{Wang2021}.
Source terms are scaled to the $^{244}$Pu global reference ($2.21\times10^3$ atoms/g). 
The depth distributions reflect the competition between impact-driven gardening and radioactive decay. A linear-scaled version of this figure is provided in the supplement (S7).}
\end{figure}

\subsection{Forward Modeling: \pu244 and Other {\em r}-Process Species} 

In order to evaluate the long-term sequestration of {\em r}-process isotopes, we forward model possible concentration profiles of \pu244{} ($t_{1/2} = 81.3$ Myr), and other {\em r}-process species produced with it. Unlike \fe60, the long half-life of \pu244{} makes it possible to track interstellar influx and gardening dynamics over timescales comparable to the age of the regolith itself. We simulate three astrophysical histories using the ADE model, see Eq.~(\ref{eq:ADE}), to determine if the vertical distribution can distinguish between episodic and steady-state delivery of \pu244, as shown in Fig.~\ref{fig:244Pu}.
All of our models utilizes a spatial grid extending from $z = 0$ to $80$ cm.

For the {\em r}-process species, 
we use the observed deep-sea ratio $(\pu244/\fe60)_{\rm deep-sea}= (4.7 \pm 5) \times 10^{-5}$ \cite{Wallner2021}
to infer the \pu244 flux from the \fe60 flux during the last 9 Myr when these have been measured.
To illustrate the result for other {\em r}-process radioisotopes, we adopt the kilonova-inspired 
KA model in ref.~\cite{Wang2023} to give production ratios to \pu244. 
The abundance ratios of \I129, \hf182, and \cm247 relative to \pu244 for H1 and H2 are taken from Table 1 of \cite{Wang2023} and Table 5 of \cite{Wang2021}, respectively.
Adopted {\em r}-process quantities appear in Table \ref{tab:flux_params}.

We consider three different flux histories to illustrate widely different
but possible {\em r}-process deposition on Earth.
We use the resulting $\Phi(t)$ in the source term (eq.~\ref{eq:source}) and defined the following histories:\\

\noindent
\textbf{History 1 (H1): Correlation with SN Pulses (solid line):} \pu244 and {\em r}-process delivery is fully coupled to the two recent supernova events ($2.3$ and $7.3$ Mya) identified in the \fe60 record, with the same Gaussian peak times and widths.
For each pulse the \pu244 fluence is derived from the \fe60 fluence: $\mathcal{F}_{244} = (\pu244/\fe60)_{\rm deep-sea} \mathcal{F}_{60}$.
%We assume an r-process fluence of $274$ and $51$ atoms/cm$^2$, respectively. 
Because the deposition is relatively recent, the signal is primarily sequestered in the upper $10$ cm, showing sharp gradients.  \\

\noindent
\textbf{History 2 (H2): Short-term Continuous Influx (dashed line):} A constant source flux of $\Phi=250\times 10^{-6}$ atoms/cm$^{2}$ is applied over the last 10 Myr. This models a steady-state accumulation of interstellar dust during the Solar System's recent passage through local debris clouds. The profile exhibits a morphology similar to the solar maturity ($\mature$).\\

\noindent
\textbf{History 3 (H3): Long-term Continuous Influx (dotted line):} The same constant \pu244 flux is applied over $80$ Myr. Over this longer duration, the advective front $v_z(t)$ and cumulative diffusion $\kappa(t)$ transport the isotope significantly deeper and more thoroughly into the column, and the integration of signal over a longer time period leads to a larger total fluence.\\

The difference between Histories 1 and 3 suggests that isotopic analysis of deep regolith samples (e.g., $z > 20$ cm) can serve as a diagnostic tool, as indicated in the left panel of Fig.~\ref{fig:244Pu}. A deep-seated \pu244{} tail would imply a persistent interstellar background, whereas a signal restricted to the upper reworking zone would support an origin tied to recent, discrete supernova events. The global empirical error envelope ($\sigma_{global} \approx 0.24$) indicates that these histories remain statistically resolvable despite the stochastic nature of impact gardening. 

H1 and H2 give similar results for \pu244.  For these cases, therefore, distinguishing between possible origins of a persistent interstellar background, e.g., multiple supernovae or kilonovae, would require complementary diagnostic tools, e.g., measurements of the density profiles of other live isotopes such as \I129{} ($t_{1/2} = 16.1$ Myr), \hf182{} ($t_{1/2} = 8.90$ Myr) or \cm247{} ($t_{1/2} = 15.6(5)$ Myr), as seen in the right panel of Fig.~\ref{fig:244Pu}. In this panel, we consider an $r$-process source that either is coincident with the iron pulses and plutonium production as in History 1 or is a separate, earlier event that is also responsible for the plutonium continuum influx in History 2. In the former, we take the $r$-process radioisotopes to be produced along with the pulses at 2.3 and 7 My at ratios consistent with the KA model of \cite{Wang2021}, while in the latter we take the $r$-process production to have occurred with the same ratios but at a much earlier time, which we take to be $t_{\rm inj} = 50 \ \rm Myr$.  This is meant to account for an earlier {\em r}-process event whose ejecta is later swept up in the Local Bubble; the production timescale consistent with our earlier work \cite{Wang2021}, and similar to but somewhat smaller than the estimated neutron star merger recurrence time in ref.~\cite{Wehmeyer2023}. 
The flux of {\em r}-process isotope $i$ which is $\Phi_{i,\rm inj}$ upon injection will decay so that later $\Phi_i(t) = \Phi_{i,\rm inj} e^{-\lambda_i (t_{\rm inj}-t)}$. 
Given the shorter halflives of the other $r$-process radioisotopes, measurements of these species relative to \pu244 can potentially distinguish between the two scenarios.

\subsection{Forward Modeling: Prospective South Polar Isotope Measurements}

{We anticipate that {\it Artemis} missions will be able gather regolith core samples that supplement and complement the information gleaned from the {\it Apollo} samples. It will be interesting to compare measurements of the \fe60 density profile near the South Pole with the {\it Apollo} measurements at mid-latitudes. If the direction to the \fe60 source is retained during transport to the Solar System~\cite{Fry2016}, the density of \fe60 deposition will depend on lunar latitude, and the comparison of {\it Apollo} measurements with South Pole data will give indications on the location of the source of the \fe60. However, the directional information may have been lost during transport from the source, e.g., due to deflection of the dust grains carrying \fe60 by interstellar magnetic fields~\cite{Fry2020}, in which case the {\it Apollo} and {\it Artemis} measurements of \fe60 should be similar.}

{As seen in the left panel of Fig.~\ref{fig:60Fe} and in Fig.~\ref{fig:244Pu}, it will be interesting to obtain regolith core samples extending to depths $\gtrsim 100$~cm, which will probe the possible deposition of live isotopes over time-scales longer than the half-life of \fe60. For example, the density profile of \pu244 relative to \fe60 may be enhanced dramatically at depth if its deposition has been occurring since before the two detected \fe60 pulses.}

\section{Discussion}

Our ADE model is consistent with the depth profiles of \fe60 measured in the {\it Apollo} samples. 
Our results are also consistent with the \fe60 signal being uniform across the lunar nearside mid-latitudes, suggesting a capture mechanism decoupled from local mineralogy. We have applied our ADE model to predict the possible depth profile of \pu244 in different deposition scenarios. If deposition of \pu244 has been taking place continuously over the past 10 million years, it may be present in the lunar regolith down to a depth ${\cal O}(10)$~cm, whereas if deposition has been continuing for 80 million years or more its depth profile may extend to a depth ${\cal O}(100)$~cm. Thus, the lunar regolith may provide insights into the history of the Solar System extending far beyond the reach of $\sim 10$ million years provided by deep-ocean deposits. As we have also discussed, searches for other live {\em r}-process isotopes such as \I129, \hf182, and \cm247 will also be informative.

Future measurements of the \fe60 signal (and those of other live isotopes) at different latitudes may provide indications of possible source locations. Samples of the regolith from near the South Pole, as will be collected by {\it Artemis} missions, will be particularly interesting in this regard, as they will provide a larger lever arm in latitude than that provided by the {\it Apollo} measurements.
Looking beyond the scope of this work, one may also adapt our ADE model, including thermal and mechanical effects of compaction, to investigate the depth distribution of lunar volatiles and solar-delivered isotopes of interest such as \isotope{He}{3}. 

{Aspects of our numerical analysis are described in S8 in the Supplemental Material.}

\begin{acknowledgments}
The development of the transport physics was supported by NASA CDAP Grant 80NSSC25K7448. We acknowledge digitized data from \cite{morris1978situ} and \cite{fimiani2016interstellar}. The work of J.E. was supported by the United Kingdom STFC Grant ST/X000753/1.
The work of B.D.F.~was supported by the NSF grant AST-2108589. 
The work of X.W. was supported by the National Natural Science Foundation of China (Grant Nos. 12494570, 12494574, 12521005), the National Key R$\&$D Program of China (2021YFA0718500), and China’s Space Origins Exploration Program.
The work of R.S.\ was supported by US Department of Energy grants DE-FG0295ER40934 and DE-SC00268442 (ENAF) and NSF award Nos.\ PHY-2020275 (N3AS) and 21-16686 (NP3M).

\end{acknowledgments}

\section*{Supplemental Material}
\subsection{S1. An ADE Model of Gardening}
\subsection{S1.1. Impact Flux and Geometric Scaling}
We assume that the impactor population bombarding the lunar surface has a cumulative size-frequency distribution (CSFD) of the form $N(>\mathcal{D}) = a \mathcal{D}^{-b}$. This gives the rate at which craters of diameter $\mathcal{D}$ or more are created per unit area. We use the flux from secondary impacts, which dominate gardening following \cite{costello2018mixing,costello2020impact,costello2021secondary} (see Table \ref{tab:model_summary} for numerical values). We define the resulting vertical evolution of the regolith using three coupled geometric processes: excavation, compaction, and burial.

\subsection{S1.2. Excavation and Compaction Zones}
\begin{figure}[b]
\includegraphics[width=0.75\linewidth]{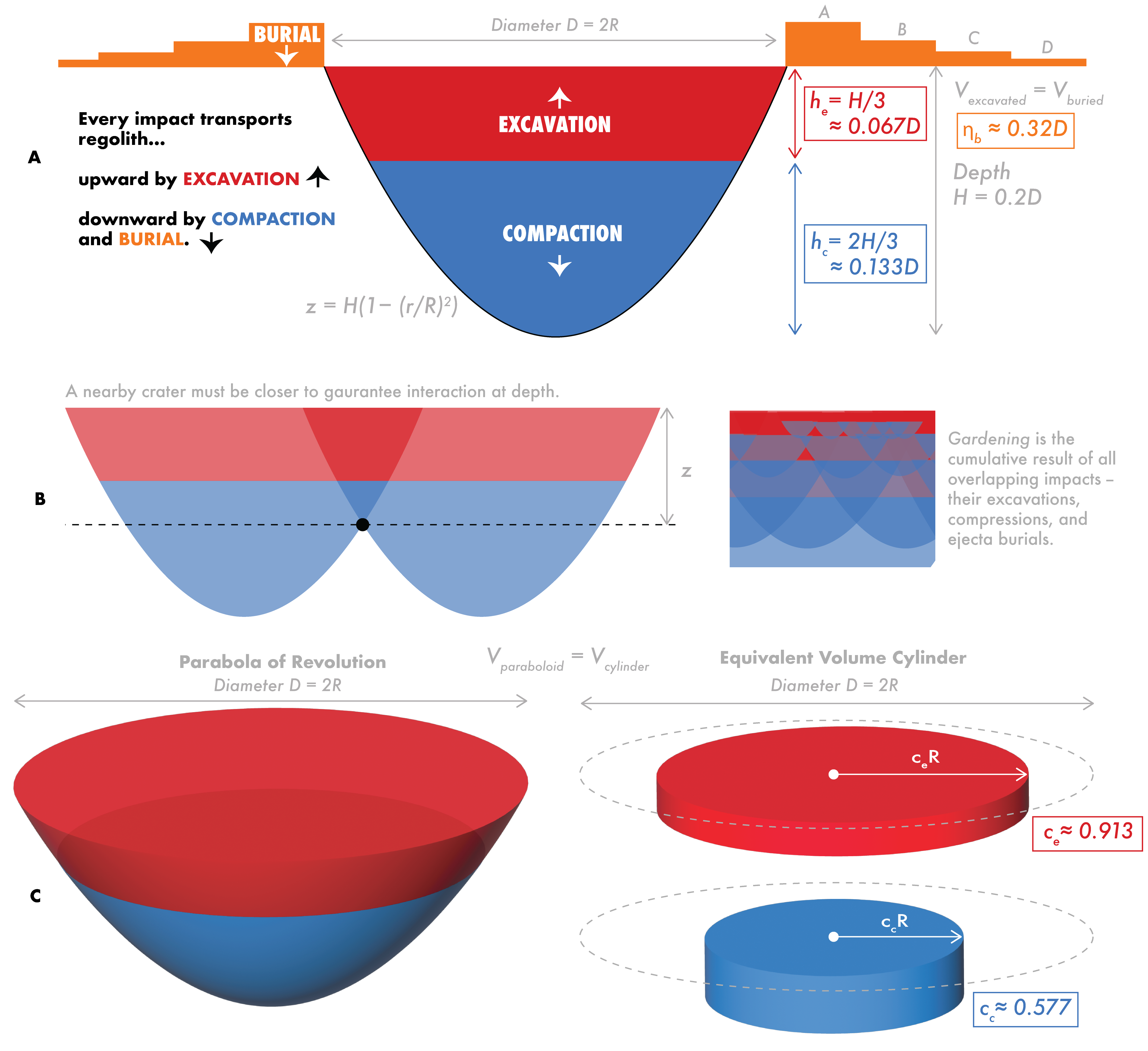}
\caption{\label{fig:Schematic} Schematic of the impact geometry underlying the gardening model.}
\end{figure}

Following \cite{costello2021secondary}, we define the characteristic gardening depth for a mechanism $i$ as:
\begin{equation}
\label{eq:AnalyticGardening}
    D_{i}(t) = 
     h_{i}\Big(\frac{  \pi a b  t c_{i}^2}{4(b-2)\lambda}\Big)^{\frac{1}{b-2}} \, ,
\end{equation}
where $h_i$ is the vertical reach and $c_i$ is the horizontal interaction scale. We assume a transient crater profile $z = H(1 - (r/R)^2)$ with a depth to diameter ratio of $H = 0.2$ (Figure \ref{fig:Schematic}).

\textbf{Excavation ($e$):} In the upper third of the crater ($\xi \in [0, 1/3]$), we assume 
\begin{equation}
    h_{e} = \frac{1}{3}H = \frac{1}{3}(0.2) \approx 0.067, \quad c_{e} = \sqrt{\frac{5}{6}} \approx 0.913 \, .
\end{equation}

\textbf{Compaction ($c$):} In the lower two-thirds of the crater ($\xi \in [1/3, 1]$), we assume
\begin{equation}
    h_{c} = \frac{2}{3}H = \frac{2}{3}(0.2) \approx 0.133, \quad c_{c} = \sqrt{\frac{1}{3}} \approx 0.577 \, .
\end{equation}
The smaller interaction radius $c_c$ reflects the parabolic tapering of the crater bowl at depth.

\subsection{S1.3. Mass Conservation and Burial}

{We use the physical model of burial developed in \cite{costello2021secondary} and illustrated in Fig.~\ref{fig:Schematic}, ensuring} volumetric consistency between material removed from the bowl of a crater and relocated as ejecta. Here, we adapt this treatment for our four discrete depositional annuli (labeled A through D). By equating a deposited volume fraction $P_{V,k}$ to the geometric volume of its receiving annulus, we derive the dimensionless burial scale $h_{b,k}$:
\begin{equation}
h_{b,k} = \frac{P_{V,k}}{18(j_k^2 - i_k^2)}
\end{equation}
where $i_k$ and $j_k$ are the inner and outer radii of the annulus in units of the transient crater radius $R$. Unlike the bulk excavation and compaction zones, the net burial is a superposition of transport from multiple spatial interaction scales $c_{b,k} = \sqrt{j_k^2 - i_k^2}$. 

By substituting these scales into the process-specific transport law, Eq. (\ref{eq:AnalyticGardening}), the total burial depth $D_B(t)$ is the sum of contributions from all ejecta annuli:
\begin{equation}
\label{eq:Burial_Full}
\begin{split}
D_B(t) &= \sum_k D_{b,k}(t) \\
&= \underbrace{ \left[ \sum_{k \in \{A, B, C, D\}} \frac{h_{b,k}}{c_{b,k}^{\frac{2}{-b+2}}} \right] }_{\text{Geometric Efficiency}} \underbrace{ \left( \frac{4(b-2)\lambda}{  \pi a b  t c_{i}^2} \right)^{\frac{1}{b-2}} }_{\text{Probabilistic Frequency}}
\end{split}
\end{equation}

This allows us to define an effective burial efficiency,
\begin{equation}
\eta_b(b) = \sum_{k \in \{A, B, C, D\}} \frac{h_{b,k}}{c_{b,k}^{\frac{2}{-b+2}}},
\end{equation}
\noindent representing the net vertical reach of the combined ejecta blanket. To evaluate $\eta_b(b)$ explicitly in terms of the fundamental parameters of the annuli, we substitute $h_{b,k} = \frac{P_{V,k}}{18(j_k^2 - i_k^2)}$ and $c_{b,k} = \sqrt{j_k^2 - i_k^2}$ into the summation:
\begin{equation}
    \eta_b(b) = \sum_{k \in \{A, B, C, D\}} \frac{\frac{P_{V,k}}{18(j_k^2 - i_k^2)}}{\left( \sqrt{j_k^2 - i_k^2} \right)^{\frac{2}{-b+2}}}
\end{equation}

Consolidating the terms in the denominator, the $(j_k^2 - i_k^2)$ factors combine as $(j_k^2 - i_k^2)^1 \cdot (j_k^2 - i_k^2)^{\frac{1}{-b+2}} = (j_k^2 - i_k^2)^{\frac{-b+3}{-b+2}}$. This yields the final closed-form solution for the effective burial efficiency:
\begin{equation}
\label{eq:EtaBurial}
    \eta_b(b) = \frac{1}{18} \sum_{k \in \{A, B, C, D\}} \frac{P_{V,k}}{\left( j_k^2 - i_k^2 \right)^{\frac{-b+3}{-b+2}}}
\end{equation}

The dimensionless inner radii $i_k$, outer radii $j_k$, and depositional volume fractions $P_{V,k}$ for each annulus are detailed in Table \ref{tab:model_summary}. For a secondary-dominated impactor population ($b \approx -4$ \cite{costello2021secondary}), evaluating this summation using these parameters yields $\eta_b \approx 0.032$, which we use as the standard burial efficiency in our advection calculations.

\begin{table*}[htb]
\centering
\caption{Model Parameters, Transport Coefficients, and Units}
\vspace{3mm}
\label{tab:model_summary}
\begin{tabular}{|l|c|l|c|l|}
\hline
\textbf{Category} & \textbf{Symbol} & \textbf{Value/Expression} & \textbf{Units} & \textbf{Description} \\ \hline
\textbf{Impact Flux} & $a$ & $1.6 \times 10^{-5}$ & cm$^{b-2}$ yr$^{-1}$ & Crater production scaling constant \\ \cline{2-5} 
 & $b$ & $4$ & -- & Crater production power-law slope \\ \cline{2-5} 
 & $\lambda$ & $4.605$ & -- & Poisson threshold ($P=99\%$) \\ \hline
\textbf{Geometry} & $h_e, c_e$ & $0.067, 0.913$ & -- & Excavation vertical and horizontal scales \\ \cline{2-5} 
 & $h_c, c_c$ & $0.133, 0.577$ & -- & Compaction vertical and horizontal scales \\ \cline{2-5} 
 & $i_k$ & $[1.0, 1.5, 2.0, 4.0]$ & $R$ & Inner radii of burial annuli $k \in \{A,B,C,D\}$ \\ \cline{2-5} 
 & $j_k$ & $[1.5, 2.0, 4.0, 11.0]$ & $R$ & Outer radii of burial annuli $k \in \{A,B,C,D\}$ \\ \cline{2-5} 
 & $P_{V,k}$ & $[0.24, 0.19, 0.05, 0.005]$ & -- & Volume fraction per annulus (S1.3) \\ \cline{2-5} 
 & $\eta_b(b)$ & $\approx 0.032$ (for $b=-4$) & -- & Effective burial efficiency (Eq. (\ref{eq:Burial_Full})) \\ \cline{2-5}  \hline
\textbf{Transport} & $v_z(t)$ & $\partial D_z / \partial t$ & cm yr$^{-1}$ & Advection (gardening) velocity \\ \cline{2-5} 
 & $\kappa(t)$ & $z_\kappa^2 / t$ & cm$^2$ yr$^{-1}$ & Diffusion coefficient \\ \cline{2-5} 
 & $\Gamma$ & $0.1077$ & -- & Gardening efficiency factor (S2) \\ \cline{2-5} 
 & $\sigma_{global}$ & $0.24$ & -- & Empirical error factor \\ \hline
\textbf{Physics} & $T_{1/2}(^{60}\text{Fe})$ & $2.62$ & Myr & $^{60}$Fe half-life \cite{NuDat} \\ \cline{2-5} 
 & $T_{1/2}(^{244}\text{Pu})$ & $80.0$ & Myr & $^{244}$Pu half-life \cite{NuDat} \\ \cline{2-5} 
 & $\mathcal{F}$ & Table \ref{tab:flux_params} & at cm$^{-2}$ & Total SN isotope fluence \\ \hline
\end{tabular}
\end{table*}

\subsection{S1.4. Net Advection and Turnover Equilibrium}

The net vertical transport depth $D_z(t)$ represents the characteristic progression of the gardening front. As shown in Eq. \ref{eq:Dz_Full}, it is the sum of process-specific depths:
\begin{equation}
\label{eq:Dz_Sum_Supp}
D_z(t) = D_B(t) + D_C(t) - D_E(t) \, .
\end{equation}

\noindent We can now derive the integrated form of Eq.~(\ref{eq:Dz_Full}) by substituting the process-specific transport depths. For compaction and excavation, we use the scaling law from Eq. (\ref{eq:AnalyticGardening}), and for burial, we use the effective efficiency $\eta_b(b)$ derived in Eq. (\ref{eq:Burial_Full}):
\begin{equation}
\begin{aligned}
D_z(t) = & \left[ \sum_k \frac{h_{b,k}}{c_{b,k}^{\frac{2}{-b+2}}} \right] \left( \frac{4(b-2)\lambda}{  \pi a b  t } \right)^{\frac{1}{b-2}} \\ 
& + \left[ \frac{h_c}{c_c^{\frac{2}{-b+2}}} \right] \left( \frac{4(b-2)\lambda}{  \pi a b  t } \right)^{\frac{1}{b-2}} \\
& - \left[ \frac{h_e}{c_e^{\frac{2}{-b+2}}} \right] \left( \frac{4(b-2)\lambda}{  \pi a b  t } \right)^{\frac{1}{b-2}} \, .
\end{aligned}
\end{equation}

\noindent Factoring out the common probabilistic frequency term yields the final analytic expression for the gardening front defined in the main text as Eq.~(\ref{eq:Dz_Full}):
\begin{equation*}
D_z(t) = \underbrace{ \left[ \eta_b(b) + \frac{h_c}{c_c^{\frac{2}{-b+2}}} - \frac{h_e}{c_e^{\frac{2}{-b+2}}} \right] }_{\text{Geometric Efficiency}} \underbrace{ \left( \frac{4(b-2)\lambda}{  \pi a b  t } \right)^{\frac{1}{b-2}} }_{\text{Probabilistic Frequency}} \, .
\end{equation*}

\noindent To define the advection of the regolith column, we take the time derivative of the gardening front to find the vertical advection velocity, $v_z(t)$:
\begin{equation}
\label{eq:velocity}
v_z(t) = \frac{\partial D_z}{\partial t} = -\frac{D_z(t)}{(-b+2)t} \, .
\end{equation}

\noindent This result completes the analytical link between the impact flux and the transport coefficients used in our model. 

At the depth scales sampled by surface operations (upper few meters), impact gardening is dominated by secondary impacts \cite{costello2021secondary}. Every primary meteorite impact generates hundreds of thousands of smaller secondaries, resulting in a power law flux of  $b \approx 4$ \cite{costello2021secondary}). For secondary impact-dominated gardening, burial and compaction dominate transport, leading to net downward advection ($v_z, D_z > 0$).

Exhumation only dominates when the production function is shallow ($b > b_{crit} \approx 3.25$), favoring the excavation of deep material over burial. This balance explains the dual nature of regolith evolution: while shallow-sloped primary impacts ($b \approx 2.7$ \cite{brown2002flux}) effectively mine the underlying bedrock to grow the total regolith column, the steep power-law flux of secondary impacts dominates the downward advection of surface-correlated material into the upper centimeter and meter depths they primarily affect.

\subsection{S2. Derivation of Transport Coefficients}

The continuum evolution of a concentration $C(z,t)$ in the lunar regolith is governed by the Advection-Diffusion Equation (ADE):
\begin{equation}
\label{eq:ADE_Supp}
\frac{\partial C}{\partial t} = \frac{\partial}{\partial z} \left( \kappa(t) \frac{\partial C}{\partial z} \right) - v_z(t) \frac{\partial C}{\partial z} - \lambda_{\text{decay}}C + S(z,t)
\end{equation}
where $\kappa(t)$ is the time-dependent diffusion coefficient and $v_z(t)$ is the vertical advection velocity. These coefficients are derived from the integrated gardening front $D_z(t)$ defined in Eq. (\ref{eq:Dz_Full}).

The vertical advection velocity represents the instantaneous rate of advance of the gardening front. We derive this by taking the time derivative of the net vertical transport depth:
\begin{equation}
v_z(t) = \frac{\partial D_z}{\partial t} = \frac{\partial}{\partial t} \left[ \eta_{\text{total}}(b) \left( \frac{4(b-2)\lambda}{  \pi a b  t } \right)^{\frac{1}{b-2}} \right]
\end{equation}

\noindent where $\eta_{\text{total}}(b)$ is the combined geometric efficiency of burial, compaction, and excavation. Because $D_z \propto t^{-1/(b+2)}$, the power-rule differentiation with respect to $t$ yields the velocity expression used in our model:
\begin{equation}
\label{eq:velocity_deriv}
v_z(t) = \frac{1}{b-2} \frac{D_z(t)}{t} \, .
\end{equation}
Defining depth $z$ as positively downward, the direction of transport is dictated by the impact population slope $b$. For a secondary-dominated population ($b \approx 4$), the leading term $-(-b+2)^{-1}$ is positive ($\approx 0.5$), resulting in the net downward advection ($v_z > 0$) of surface-delivered materials.

While $D_z(t)$ and $v_z(t)$ describe the bulk advective progression of the regolith column, impact gardening is fundamentally stochastic. The regolith is continuously churned by numerous smaller craters that do not independently drive the net advective front but dominate local mixing. We capture this stochastic mixing as a diffusive process, where the time-dependent diffusion coefficient $\kappa(t)$ scales with the square of a characteristic mixing length $z_\kappa(t)$:
\begin{equation}
\kappa(t) = \frac{z_\kappa(t)^2}{t} \, .
\end{equation}
The scale of this local mixing is physically coupled to the net transport depth $D_z(t)$ via two dimensionless geometric properties of an average impact event: the total vertical reach ($H$) and the gardening efficiency ($\Gamma$). The total vertical reach characterizes the combined vertical scale of crater formation. We partition this total depth into the excavation zone ($h_e \approx 0.067$) and the underlying sub-crater compaction zone ($h_c \approx 0.133$). Their sum precisely corresponds to the transient depth-to-diameter ratio ($d/D \approx 0.2$ \cite{melosh1989impact}) of simple lunar craters, such that $H = h_e + h_c = 0.2$.

The gardening efficiency ($\Gamma$) captures the spatial variance of impact-driven vertical displacements across all depositional and formational zones. We derive $\Gamma$ by taking the root-sum-square of the individual vertical geometric scales ($h_e$, $h_c$, and the effective burial efficiency $\eta_b$), weighted by the mean horizontal interaction fraction $\bar{c} = 1/\sqrt{2} \approx 0.707$:
\begin{equation}
\Gamma = \bar{c} \sqrt{h_e^2 + h_c^2 + \eta_b(b)^2} \, .
\end{equation}
For our secondary-dominated parameters ($h_e = 0.067$, $h_c = 0.133$, and $\eta_b(-4) \approx 0.032$), this yields a dimensionless gardening efficiency of $\Gamma \approx 0.108$. The characteristic mixing length $z_\kappa(t)$ is then defined by the ratio of this spatial variance to the total vertical reach, scaled by the current advection depth:
\begin{equation}
z_\kappa(t) = \frac{\Gamma}{H} D_z(t) \approx \frac{0.108}{0.2} D_z(t) = 0.54 D_z(t) \, .
\end{equation}
Substituting this relationship into our definition of $\kappa(t)$ yields the final analytical expression for the time-varying diffusion coefficient:
\begin{equation}
\label{eq:diffusion_deriv}
\kappa(t) = \left( \frac{\Gamma}{H} \right)^2 \frac{D_z(t)^2}{t} \, .
\end{equation}
\noindent This formulation physically links the rate of diffusive spreading to the advective progression of the regolith column, with both transport coefficients ($\kappa, v_z$) naturally decaying as $t$ increases and the gardening front moves deeper into the regolith.

\subsection{S3. Concentration with Depth}

The continuum evolution of a concentration $C(z,t)$ in the lunar regolith is governed by the conservation of mass. We evaluate the rate of change of concentration at any depth $z$ by taking the negative divergence of the vertical mass flux $J(z,t)$, combined with internal sinks (decay) and sources:
\begin{equation}
\frac{\partial C}{\partial t} = -\frac{\partial J}{\partial z} - \lambda_{\text{decay}} C + S(z,t) \, 
\end{equation}
where $\lambda_{\text{decay}} = 1/\tau$ is the radioactive decay rate (for stable maturity metrics, $\lambda_{\text{decay}} = 0$) and $S(z,t)$ represents the spatiotemporal source term, combining continuous surface delivery, episodic deposition, and in-situ cosmogenic production. 

The total vertical mass flux $J(z,t)$ consists of two distinct physical transport mechanisms driven by impact gardening:
\begin{enumerate}
    \item \textbf{Advective Flux ($J_{\text{adv}}$):} The bulk, directed transport of the regolith column due to the net progression of the gardening front. This is governed by the time-varying velocity $v_z(t)$, resulting in $J_{\text{adv}} = v_z(t) C$.
    \item \textbf{Diffusive Flux ($J_{\text{diff}}$):} The stochastic, bidirectional mixing caused by numerous smaller impacts above the main advective front. This flux is proportional to the concentration gradient, such that $J_{\text{diff}} = -\kappa(t) \frac{\partial C}{\partial z}$.
\end{enumerate}

\noindent Substituting these flux components ($J = J_{\text{adv}} + J_{\text{diff}}$) into the mass conservation equation yields the general form of the Advection-Diffusion Equation (ADE) presented in the main text as Eq.~(\ref{eq:ADE}):
\begin{equation}
\frac{\partial C}{\partial t} = \frac{\partial}{\partial z} \left( \kappa(t) \frac{\partial C}{\partial z} \right) - \frac{\partial}{\partial z} \left( v_z(t) C \right) - \lambda_{\text{decay}} C + S(z,t) \, .
\end{equation}

\noindent Because our transport coefficients $\kappa(t)$ and $v_z(t)$ are derived from the integrated impact history (Eqs. (\ref{eq:velocity}) and (\ref{eq:diffusion})), they are strictly time-dependent and spatially uniform across the defined regolith column (i.e., $\frac{\partial \kappa}{\partial z} = 0$ and $\frac{\partial v_z}{\partial z} = 0$). This allows us to pull the coefficients outside the spatial derivatives, recovering the final working form of the equation used to calculate our depth profiles:
\begin{equation}
\label{eq:ADE_Working}
\frac{\partial C}{\partial t} = \kappa(t) \frac{\partial^2 C}{\partial z^2} - v_z(t) \frac{\partial C}{\partial z} - \lambda_{\text{decay}}C + S(z,t) \, .
\end{equation}
When solving this equation numerically, we treat the vacuum-regolith interface ($z=0$) as a flux boundary.

\subsection{S3.1. Continuous Influx (Maturity)}
For the calculation of $\mature$, we assume $\lambda = 0$ and $S(z,t) = \frac{\phi_{m}}{\rho(z)}\delta(z)$, where $\phi_{m}$ is the constant maturity influx rate and $\rho(z)$ is the density. In this context, $C(z,t)$ represents the accumulated maturity (normalized to FeO) resulting from surface exposure.

\subsection{S3.2. Sporadic Pulses and Radionuclides:}
The model generalizes to any radionuclide. For an isotope with a half-life $t_{1/2}$, the decay constant is $\lambda = \ln(2)/t_{1/2}$. The total source $S(z,t)$ is normalized by the abundance of a target element (e.g., wt\% FeO for $^{60}\text{Fe}$) as follows:

\begin{equation}
S(z,t) = \frac{\Phi(t)}{\rho(z)}\delta(z) + P_0 \cdot \exp\left( -\frac{\xi(z)}{\Lambda} \right) \, ,
\end{equation}
where $P_0$ is the normalized surface production rate by galactic cosmic rays (GCR), which attenuates with mass-depth $\xi(z) = \int_0^z \rho(z') dz'$, and $\Lambda$ is the attenuation length.

The external surface flux $\Phi(t)$ for radionuclides such as $^{60}$Fe is modeled as a summation of discrete events, each represented by a Gaussian distribution in time:

\begin{equation}
\Phi(t) = \phi_{bg} + \sum_{i} \frac{\mathcal{F}_i}{\sigma_i \sqrt{2\pi}} \exp\left( -\frac{(\tau - t_i)^2}{2\sigma_i^2} \right) \, ,
\end{equation}

\noindent where $\tau$ is the look-back time, $\mathcal{F}_i$ is the event magnitude (fluence) or background flux ($\phi_{bg}$), $\sigma_i$ is the temporal width, and $t_i$ is the time of the peak flux.

\subsection{S3.3. Isotope Source Parameters}

\begin{figure}[t]
\includegraphics[width=0.75\linewidth]{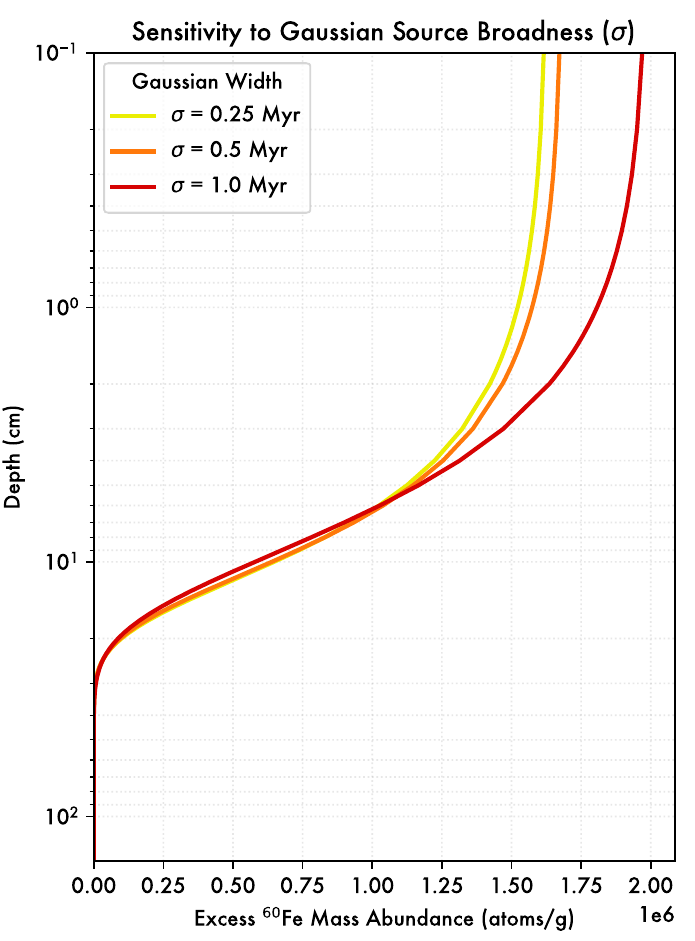}
\caption{{Illustration of the sensitivity of the calculated \fe60 density profile to the Gaussian width assumed for the more recent pulse pulse.}}
\label{fig:width}
\end{figure}

The age and width of the more recent live-isotope pulse are taken from~\cite{Ertel2023}, and its \fe60  fluence is taken from~\cite{Wallner2021}. The parameters of the earlier \fe60 pulse are taken from the data of~\cite{Wallner2021}, using the updated analysis described in~\cite{Koll2023}. The \pu244 fluences and continuous flux are taken from~\cite{Wallner2021}. {
The isotope source scenarios and flux parameters that we consider are shown in Table~\ref{tab:flux_params}}. Fig.~\ref{fig:width} shows the sensitivity of the the calculated \fe60 density profile to the Gaussian width assumed for the more recent pulse. The variation in the density profile is $\sim 20$\% at depths $\lesssim 1$~cm and much smaller in the tail at depths $\gtrsim 10$~cm.

\begin{table*}[htb]
\centering
\caption{Isotope Source Histories and Flux Parameters}
\vspace{3mm}
\label{tab:flux_params}
\begin{tabular}{l l c c c}
\hline
\textbf{Isotope} & \textbf{Scenario} & \textbf{Peak ($t_i$)} & \textbf{Width ($\sigma_i$)} & \textbf{Fluence/Flux ($\mathcal{F}_i$/$\phi_{bg}$)} \\
\hline
$^{60}\text{Fe}$ & SN Pulses & 2.3 Mya & 0.47 Myr & $3.6 \pm 0.2 \times 10^7$ at/cm$^2$ \\
                 &           & 7.3 Mya & 0.25 Myr & $1.1 \pm 0.2 \times 10^7$ at/cm$^2$ \\
\hline
$^{244}\text{Pu}$ & H1: Sporadic & 2.3 Mya & 0.47 Myr & $1.7 \times 10^3 \ {\rm at/cm^2}$ \\
                  &              & 7.3 Mya & 0.25 Myr & $5.0 \times 10^2 \ {\rm at/cm^2}$ \\
\cline{2-5}
                  & H2 and H3:  Continuous & -- & -- & $246 \ {\rm at \, cm^{-2} \, Myr^{-1}}$  \\
\hline
\end{tabular}
\end{table*}

\subsection{S.3.4. Cosmogenic Production and Gardening Steady State } \label{sec:cosmo_steady}

To evaluate the background concentration of radionuclides against which episodic supernova signals must be identified, we model the steady-state profile resulting from in-situ production by galactic cosmic rays (GCR) and impact-driven redistribution. This background is established by the balance between depth-dependent production, radioactive decay, and vertical transport.

\subsubsection{Depth-Dependent Production Rate}
The volumetric production rate $S_{\text{cosmo}}(z,t)$ (atoms cm$^{-3}$ yr$^{-1}$) is governed by the attenuation of the cosmic ray flux as it penetrates the regolith. We model this as a function of the cumulative mass depth $\xi(z)$ (g cm$^{-2}$) to account for the depth-varying density of the soil \cite{reedy1985model, masarik1994effects}:
\begin{equation}
S_{\text{cosmo}}(z,t) = P_0 \cdot \rho(z) \cdot \exp\left( -\frac{\xi(z)}{\Lambda} \right)
\end{equation}
where $P_0$ is the surface production rate per unit mass (normalized to $\approx 0.039$ atoms g$^{-1}$ yr$^{-1}$ for \fe60, based on the cosmogenic background flux estimated in \cite{fimiani2016interstellar, Zwickel2026}), and $\Lambda = 160$ g cm$^{-2}$ selected as the attenuation length for galactic cosmic ray (GCR) spallation and capture reactions \cite{reedy1985model,masarik1994effects}.

\subsubsection{Steady-State and  Equilibrium}
The evolution of the cosmogenic component is integrated using the unified Advection-Diffusion Eq.~(\ref{eq:ADE}). Unlike the episodic supernova pulses, which are introduced as surface delta functions $\delta(z)$, the cosmogenic source is continuous and distributed throughout the regolith column.

In simulations we integrate the system for a simulation time ($T_{\text{sim}} = 20\text{--}50$ Myr) significantly longer than the mean life of the isotope ($\tau(^{60}\text{Fe}) \approx 3.78$ Myr) so that profile reaches a secular equilibrium where the total inventory is governed by $\int S(z) dz = \lambda \int C(z) dz$, and the gardening transport parameters ($v_z, \kappa$) have reached their long-term average behavior. 

\subsubsection{Normalization to Site Chemistry}
Empirical measurements of \fe60 are typically reported as activities relative to nickel (\fe60/Ni). However, because cosmogenic production also occurs on iron targets, the resulting concentration is sensitive to the local weight fraction of iron oxide ($w_{\text{FeO}}$) at different {\it Apollo} landing sites. To facilitate a direct comparison between model outputs and diverse site data (e.g., {\it Apollo} 12 basalts vs. {\it Apollo} 16 highlands), we normalize the concentrations:
\begin{equation}
C_{\text{norm}}(z) = \frac{C(z)}{\rho(z) \cdot w_{\text{FeO}}}
\end{equation}
This normalization expresses the concentration in units of atoms g$^{-1}$ per 1 wt\% FeO. By doing so, the cosmogenic floor becomes a site-independent baseline, allowing for the clear identification of excess \fe60 delivered by interstellar pulses regardless of local major-element chemistry.

\subsection{S4. Empirical Uncertainty and Model Stochasticity}

To account for the inherent stochasticity of lunar regolith gardening, we define an empirical uncertainty envelope based on the characteristic deviation of observed maturity profiles from mean-field model predictions. We utilize the digitized $\mature$ depth profiles from \cite{morris1978situ} across the {\it Apollo} 15, 16, and 17 landing sites as proxies for long-term gardening effects.

The empirical error calculation follows a three-step normalization and aggregation procedure:
\begin{enumerate}
    \item \textbf{Residual Calculation:} For each reference core $s$, the mean-field transport model $M(z)$ was solved for the specific exposure age of the site and interpolated to the exact measurement depths $z_i$ of the experimental data $D(z_i)$. Residuals $\epsilon_{i,s}$ were calculated on a normalized scale relative to the profile maximum:
    \begin{equation}
        \epsilon_{i,s} = \frac{D(z_{i,s})}{\max(D_s)} - \frac{M(z_{i,s})}{\max(M_s)} \, .
    \end{equation}
    
    \item \textbf{Global Variance:} A site-agnostic standard deviation, $\sigma_{global}$, was derived from the combined residuals of all reference profiles. This parameter captures the universal noise introduced by discrete impact events that deviate from the continuous diffusion-advection approximation:
    \begin{equation}
        \sigma_{global} = \sqrt{\frac{1}{N} \sum_{i,s} \epsilon_{i,s}^2} \approx 0.24 \, .
    \end{equation}
    
    \item \textbf{Model Application:} The uncertainty in forward-modeled species (e.g.,  \fe60 and \pu244)) is expressed as a $1\sigma$ envelope. Because the gardening intensity scales with the total concentration, the absolute error is defined proportional to the peak concentration $C_{max}$ of the resulting profile:
    \begin{equation}
        C_{err}(z) = C_{model}(z) \pm (\sigma_{global} \times C_{max}) \, .
    \end{equation}
\end{enumerate}

\noindent
This approach ensures that the model provides a statistically realistic representation of the vertical distribution while acknowledging that individual core samples are single realizations of a highly variable impact history.

\subsection{S5. Space Weathering and \fe60}

{Pitting and spall by micro-craters and amorphization of the crystal structure by solar wind irradiation systematically alter the external microstructure of mineral grains \cite[e.g., ][]{pieters2016space, denevi2023space, cao2025nature}. A primary consequence of this alteration is the precipitation of nanophase metallic iron (np-Fe$^0 \equiv I_s$) within regolith grain rims, which determines the measured $\mature$ maturity index (Fig.~\ref{fig:Morris}) and is known to concentrate in the finest, highly mature fractions of the soil closest to the lunar surface \cite{wiesli2003space, pieters2016space}. Because supernova-derived \fe60 is delivered externally via interstellar dust, it may preferentially accumulate on these damaged, highly reactive grain exteriors \cite{Zwickel2026}.}

{Conversely, the in-situ cosmogenic \fe60 background is produced primarily within the grain interior through cosmic-ray spallation on stable Ni \cite{Zwickel2026}. Space weathering actively drives chemical and isotopic fractionation; impact-induced evaporation from continuous micrometeorite bombardment causes the preferential loss of lighter isotopes and elements like Ni, a mechanism recently evidenced in young lunar drill cores \cite{li2025significantly}. Therefore, the extensive structural damage and the generation of chemically reactive np-Fe$^0$ rims in the finest mature regolith provide enhanced capacity for capturing recent exogenic \fe60, while the simultaneous weathering-induced loss of cosmogenic \fe60 and Ni limits in-situ cosmogenic production, ultimately revealing a pronounced supernova signature.} 

\subsection{S6. Regolith Density and Volumetric Abundance}

To facilitate a direct comparison between laboratory-measured specific activities—typically reported as activity per unit mass of nickel—and our numerical transport model, we transformed the observed abundances into a volumetric number density, $n(z)$, with units of $\text{atoms/cm}^3$. A critical factor in this transformation is the depth-dependency of the lunar bulk density, $\rho(z)$. The lunar regolith is characterized by a ``fluffy" surface layer that compacts with depth due to overburden pressure. 

Following the empirical results from the Lunar Reconnaissance Orbiter (LRO) Diviner Thermal Radiometer \cite{vasavada2012lunar, hayne2017global}, we modeled the regolith density profile using the exponential H-parameter formula:

\begin{equation}
\rho(z) = \rho_d - (\rho_d - \rho_s) e^{-z/H} \, .
\end{equation}
where $\rho_s = 1.3 \text{ g/cm}^3$ represents the surface density, $\rho_d = 1.9 \text{ g/cm}^3$ is the compacted deep density, and $H = 7.0 \text{ cm}$ is the characteristic scaling depth. The number density of a specific supernova-derived isotope at depth $z$ is then derived using:

\begin{equation}
n(z) = \left( \frac{A(z) \cdot C_{\text{Ni}}(z)}{\lambda_{decay}} \right) \rho(z) \cdot \Psi \, .
\end{equation}
In this expression, $A(z)$ is the measured activity (e.g., dpm per kg-Ni), $C_{\text{Ni}}(z)$ is the clean nickel concentration ($\mu\text{g/g}$), $\lambda_{decay}$ is the isotopic decay constant, and $\Psi$ is a composite conversion factor ($10^{-9}$) required to reconcile mass and volume scales. 

This depth-dependent correction accounts for the increased density of the lower regolith, scaling the soil's capacity for supernova fluences and preventing artificially inflated number densities near the surface.

\subsection{S7. Numerical Implementation}

All numerical modeling, data processing, and visualization were implemented using the Python programming language. The core mathematical operations and array manipulations were handled using the \texttt{NumPy} library. 

To solve the one-dimensional advection-diffusion equation governing the vertical transport of isotopes in the lunar regolith, we discretized the spatial derivatives along a non-uniform 1D depth grid using \texttt{NumPy}, with high resolution (0.1 cm spacing) in the top 2 cm to capture sharp surface boundary dynamics, relaxing to 1.0 cm spacing down to a depth of 5 m. 

The advective velocity ($v_z$) in our model emerges analytically from the time-derivative of the net transport depth, representing a competition between downward drivers (burial and compaction) and upward exhumation. Because lunar regolith gardening is dominated by a steep size-frequency distribution of secondary impactors ($b \approx 4$), this derivative mathematically dictates a net downward advection of material in the upper meter ($v_z > 0$). To respect this physical directionality and stably model the downward advective front, we implemented a first-order upwind differencing scheme. By calculating the advective flux based strictly on the upstream (shallower) regolith layers, this method prevents numerical dispersion.

The diffusion term, representing the isotropic mixing of regolith by frequent, small impacts, was discretized using standard second-order central differencing. Cosmogenic production and exogenic inputs (astrophysical deposition events) were treated as explicit source terms. The discrete astrophysical events were introduced dynamically at the surface boundary as a time-dependent flux modeled as Gaussian pulses in time.

The resulting system of coupled ordinary differential equations was integrated over the multi-million-year simulation time using the \texttt{solve\_ivp} function from the \texttt{SciPy} library. We selected the \texttt{Radau} method (an implicit Runge-Kutta scheme) to handle the stiffness of the system caused by the stark contrast between continuous, slow background processes and highly transient astrophysical pulses. To guarantee that the adaptive time-stepper did not skip over these geologically brief deposition events, we enforced a strict maximum step size of $10^5$ years. Combined with strict absolute ($10^{-7}$) and relative ($10^{-4}$) tolerances, this configuration ensures mass conservation and temporally resolves the exact onset and decay of the modeled $^{60}$Fe pulses.

The ingestion of observational {\it Apollo} sample datasets and normalization to local FeO weight percentages, and statistical handling of were managed using the \texttt{Pandas} library. All graphical outputs, error bound calculations, and model-data profile visualizations were subsequently generated using \texttt{Matplotlib}.

\subsection{S8. Linear Version of Plot for Sample Strategy}
\begin{figure}[t]
\includegraphics[width=\linewidth]{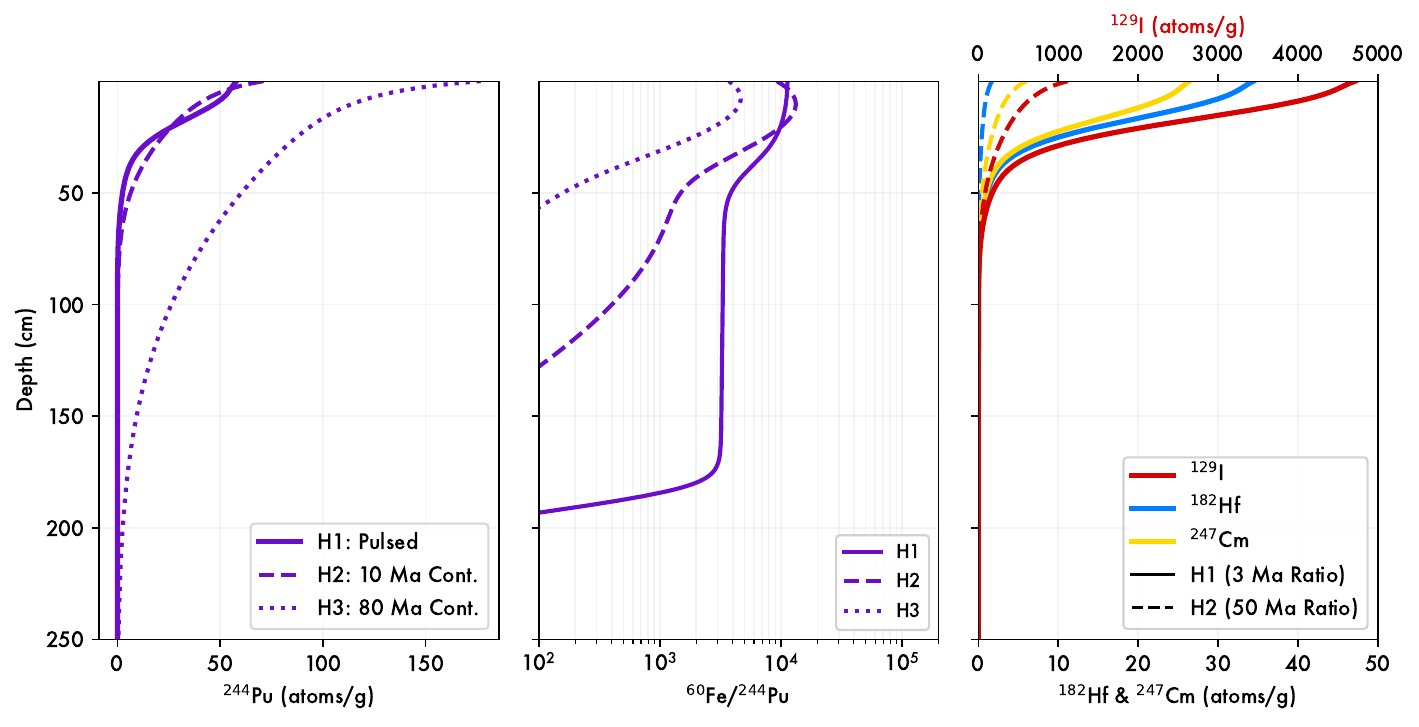}
\caption{\label{fig:244Pulin} A linear-scale version of Figure \ref{fig:244Pu} in the main text.}
\end{figure}

We provide a linear-scale version of the depth-distribution predictions for isotopes of interest to support future sample collection planning and analysis (Figure \ref{fig:244Pulin}). 

\bibliography{Gardening}

\end{document}